# Ontology-based systematic classification and analysis of coronaviruses, hosts, and host-coronavirus interactions towards deep understanding of COVID-19

Hong Yu[1,2,*], Li Li[3,*], Hsin-hui Huang[4,5*], Yang Wang[1,2,4*], Yingtong Liu[4], Edison Ong[4], Anthony Huffman[4], Tao Zeng[6], Jingsong Zhang[6], Pengpai Li[7], Zhiping Liu[7], Xiangyan Zhang[1,2], Xianwei Ye[1,2], Samuel K. Handelman[4], Gerry Higgins[4], Gilbert S. Omenn[4], Brian Athey[4], Junguk Hur[8,ξ], Luonan Chen[6,ξ], Yongqun He[4,ξ]

1. Department of Respiratory and Critical Care Medicine, Guizhou Province People's Hospital and NHC Key Laboratory of Immunological Diseases, People's Hospital of Guizhou University, Guiyang, Guizhou 550002, China.
2. Department of Basic Medicine, Guizhou University Medical College, Guiyang, Guizhou 550025, China.
3. Department of Genetics, Harvard Medical School, Boston, MA 02115, USA.
4. University of Michigan Medical School, Ann Arbor, MI 48109, USA
5. Department of Biotechnology and Laboratory Science in Medicine, National Yang-Ming University, Taipei, Taiwan.
6. Key Laboratory of Systems Biology, Center for Excellence in Molecular Cell Science, Shanghai Institute of Biochemistry and Cell Biology, Chinese Academy of Sciences, Shanghai 200031, China.
7. Center of Intelligent Medicine, School of Control Science and Engineering, Shandong University, Jinan, Shandong 250061, China
8. University of North Dakota School of Medicine and Health Sciences, Grand Forks, ND 58203, USA.

* Co-first authors; ξ Co-corresponding authors

**ABSTRACT**

Given the existing COVID-19 pandemic worldwide, it is critical to systematically study the interactions between hosts and coronaviruses including SARS-Cov, MERS-Cov, and SARS-CoV-2 (cause of COVID-19). We first created four host-pathogen interaction (HPI)-Outcome postulates, and generated a HPI-Outcome model as the basis for understanding host-coronavirus interactions (HCI) and their relations with the disease outcomes. We hypothesized that ontology can be used as an integrative platform to classify and analyze HCI and disease outcomes. Accordingly, we annotated and categorized different coronaviruses, hosts, and phenotypes using ontologies and identified their relations. Various COVID-19 phenotypes are hypothesized to be caused by the backend HCI mechanisms. To further identify the causal HCI-outcome relations, we collected 35 experimentally-verified HCI protein-protein interactions (PPIs), and applied literature mining to identify additional host PPIs in response to coronavirus infections. The results were formulated in a logical ontology representation for integrative HCI-outcome understanding. Using known PPIs as baits, we also developed and applied a domain-inferred prediction method to predict new PPIs and identified their pathological targets on multiple organs. Overall, our proposed ontology-based integrative framework combined with computational predictions can be used to support fundamental understanding of the intricate interactions between human patients and coronaviruses (including SARS-CoV-2) and their association with various disease outcomes.

Human coronaviruses pose a huge threat to global public health. The COVID-19 outbreak has led to significant morbidity, mortality and economic damage worldwide [1]. As of May 24, 2020, WHO reported over 5.2 million confirmed cases and over 337,000 deaths globally. COVID-19 is caused by SARS-CoV-2, a positive sense RNA virus in the coronavirus family (*Coronaviridae*) [2,3]. COVID-19 is the third well-attested outbreak of zoonotic (infecting across species) coronaviruses in the last 18 years. The severe acute respiratory syndrome (SARS) emerged in China in 2002, and the SARS outbreak lasted for 8 months and caused in 8,098 confirmed human cases in 29 countries and 774 deaths[4,5]. In 2012, the Middle East respiratory syndrome coronavirus (MERS-CoV) was isolated in Saudi Arabia from the sputum of a male patient who died from acute pneumonia and renal failure [6]. The MERS-CoV outbreaks resulted in 2,260 cases across 27 countries and 803 deaths[7,8].

Coronaviruses are important pathogens for humans and other vertebrates [9]. Prevalent coronaviruses cause agriculturally important diseases in animals and a cold-like illness in humans [10]. Examples of coronavirus-induced animal diseases are avian infectious bronchitis virus (IBV), transmissible gastroenteritis virus (TGEV), porcine epidemic diarrhea virus (PEDV), and swine acute diarrhea syndrome coronavirus (SADS-CoV). A 2015 study estimated that IBV caused $3,567 in losses per 1000 birds on poultry farms in Brazil [11]. Prevalent human coronavirus strains, including human coronavirus 229E (HCoV-229E), HCoV-OC43, HCoV-NL63, and HKU1, continuously circulate in human populations and cause mild illness in children and adults around the world [12].

These crucial differences between SARS-CoV-2 and more benign coronaviruses are a function of specific adaptations that SARS-CoV-2 has made to the human host. For example, the immune-evasive character of SARS-CoV-2 enables the COVID-19 pandemic to spread from asymptomatic individuals [13], because symptoms are primarily produced by the host immune response. This spread from asymptomatic carriers is a key portion of COVID-19's rapid spread. Ongoing inquiry into these specific adaptations is hampered by a lack of structured integrative resources with standard data and metadata formats.

Deeper study of the interactions between hosts (human and others) and coronaviruses (SARS-CoV-2, SARS-CoV, MERS-CoV and others) would help our understanding of COVID-19 mechanisms and rational design of vaccines and drugs against COVID-19. However, to achieve this goal, two major bottlenecks exist. First, although intensive research has generated large amounts of data, the data are multidimensional, heterogeneous, disintegrated and non-interoperable. Such data cannot be efficiently interpreted by computers, preventing computer-assisted reasoning and analysis. The second bottleneck is the lack of efficient computational algorithms and bioinformatics tools to integrate and analyze the heterogeneous data and knowledge into deep scientific understanding. This bottleneck slows the integration of host-coronavirus interaction (HCI) data, discovery of HCI mechanisms, and new applications. Ontologies can solve these problems in analysis by integrating big data in a computer-interpretable way.

In the informatics field, an ontology is a human- and computer-interpretable set of terms and relations that represent entities in a specific domain and how they relate to each other. Ontology has played an important role in standard knowledge and data representation, integration, sharing, and analysis. A major field of artificial intelligence is knowledge representation and reasoning ($KR^2$, KR&R), and ontology is a foundation of the KR&R. The Coronavirus Infectious Disease Ontology (CIDO) is a new community-based biomedical ontology in the domain of coronavirus diseases (https://github.com/cido-ontology/cido) [14]. To better study the HCI mechanisms to explain various COVID-19 phenomena, we propose to use CIDO as a platform for standard and computer-understandable representation of the host-coronavirus interactions and their influences on patient outcomes.

In this manuscript, we introduce a set of "HPI-Outcome postulates", laying out a set of basic rules for understanding host-pathogen interactions (HPIs) and their roles in the disease outcomes. Then, we propose a model to study the relations between host-coronavirus interactions (HCIs) and coronavirus disease outcomes. We used our ontology-based strategy to annotate and classify human coronaviruses, hosts, and experimentally verified host-coronavirus interactions (HCIs) obtained from literature mining and bioinformatics studies. Using these annotations, we proposed an ontological model of the interactions between SARS-CoV-2 and human to explain how the human-SARS-CoV-2 interactions would induce different disease outcomes such as hypertension and cytokine storm. Furthermore, using known PPIs as baits, we developed a computational method to predict new PPIs and their associations with multi-organ disease phenotypes caused by SARS-CoV-2.

**Methods**

**Manual literature mining of virulence factors and protein-protein interacions.** From the literature, we manually annotated and identified coronavirus virulence factors and how host genes/proteins interact with coronaviruses. Experimentally confirmed coronavirus virulence factors were added to the Victors, a manually curated web-based database and analysis system[15].

**Ignet literature mining analysis**. The Ignet interaction network tool (http://ignet.hegroup.org/), powered by our in-house SciMiner [16] literature mining tool, was used to mine gene-gene interactions in the specific domain of coronaviruses.

**STRING bioinformatic analysis of PPIs.** Known protein-protein interactions among the host genes were collected from the STRING database with a medium confidence score of 0.4.

**Ontology-based representation and analysis.** Coronaviruses and their prevalent host species were mapped to NCBITaxon taxonomy ontology [17] and Human Phenotype Ontology (HPO) [18]. The disease phenotype outcomes, including different symptoms and signs, were mapped to HPO. With these mapped terms, Ontofox [19] was used to generate subsets (or slimes) of the NCBITaxon or HPO ontologies.

**Prediction of host-coronavirus PPIs.** We used a "domain-inferred prediction" method to infer inter-species PPIs. Supported by the conservation and importance of domains in proteins [20], such domain-domain interaction (DDI) infers the potential interaction between two proteins if one domain in a pathogen protein A interacted with another domain in a host protein B.

**Visualization of host-coronavirus PPIs.** Cytoscape[21] was utilized for visualization of the coronavirus-human PPIs.

**Results**

**"HPI-Outcome postulates" and its associated model for deep COVID-19 study**

Koch's postulates[22-24] have long provided criteria designed to establish a causal relationship between exposure to a microbe and emergence of a disease. The four criteria of Koch's postulates are: (i) The microbe must be found in all organisms suffering from the disease, but should not be found in healthy organisms. (2) The microbe must be isolated from a diseased organism and grown in pure culture. (3) The cultured microbe should cause the same disease when introduced into a healthy organism. (4) The microbe must be reisolated from the inoculated, diseased experimental host and be identical to the original causative agent.

If we assume that we know the causal relation between a microbe and a disease, we are then asked to explain how the host-pathogen interaction (HPI) is causally associated with an outcome. Such outcomes can include hard endpoints (*i.e.* death, kidney failure) or intermediate phenotypic outcomes such as fever, reduced blood oxygen. In this case, we may propose a set of HPI-Outcome postulates that can explain how HPIs are linked to individual disease outcomes. Like Koch's postulates, our HPI-Outcome postulates include four criteria:

(i) **HPI motivation**: The host (or pathogen) reacts to the pathogen (or host) to achieve their best possible outcome. Such phenomena explain the root motivation of the HPI and their associated outcomes.

(ii) **HPI-outcome causality**: The host or pathogen outcomes are determined by the HPIs at the molecular and cellular level. As a result, the host or pathogen's genetic and phenotypic profiles (such as the direct and co-morbid susceptibility factors of the host) may affect such HPIs and thus disease outcomes.

(iii) **HPI role**: The HPIs involves a network of molecular interactions at the interface of host and pathogen or within either the host and pathogen, and different molecules and interactions play different roles in determining disease outcomes.

(iv) **HPI interruption**: An internal or external interruption of one or more H-P interactions critical to the HPI-outcome network may change the outcome. The external interruption can arise from drug or vaccine administration.

The HPI-outcome postulates provide a set of very basic rules to guide our study of specific HPI mechanisms to explain disease outcomes. The first postulate explains the root motivation of the HPI behaviors and disease outcomes. Note that pathogens sometimes induce maladaptive responses in their hosts. For example, in the case of coronavirus, the evasion or hijacking of the host immune response is crucial to viral spread [25]. Hosts have evolved sophisticated mechanisms to detect and fight against viral infection, and immune evasion is a strategy by the virus to avoid a range of host responses. At some stage, the host will successfully mount an immune response. However, if immune evasion delays this immune response too long, the viral infection may be too widespread and the immune response itself can kill the host/patient [26].

The other postulates are also critical for HPI study. For example, the 2$^{nd}$ postulate explains why comorbid conditions may affect HPIs and thus outcomes. While in vitro or labortary animal studies approximate *in vivo* human mechanisms, such studies may not reflect HPI-outcome causal linkages in the relevant host (e.g., human) because important HPIs leading to relevant outcomes may differ between hosts. According to the 3$^{rd}$ postulate, each specific HPI initiates a dynamic interaction network and each interaction has its role in the network. The HPI may change a cascade of molecular interactions inside the host or pathogen. The roles of different molecules and interactions in the network should be carefully studied. According to the 4$^{th}$ postulate, the HPI interruption may be performed by the host or pathogen naturally or can be initiated by the administration of a drug or vaccine.

Based on the HPI-Outcome postulates, we lay out the integrative host-coronavirus interaction framework as shown in Figure 1. Basically, the framework includes three major processes from the viral side and three major processes from the host side. For the coronaviruses, crucial processes include: viral entry to host cell, viral replication, and viral release from the infected cell. For the host, three responses exist: the naïve response, the innate response, and the adaptive response. Initially, the host response to the coronavirus entry through a receptor-ligand interaction is naïve since the host does not realize the danger and react to the virus by a routine procedure. At this naïve host response stage, neither signalling molecules nor selected immune cell populations are yet present to coordinate an immune response. As signalling molecules are produced by infected cells (*e.g.* cytokines, interferons), the host will show a delayed innate responses to coronavirus infection. The innate immune response may or may not become effective and strong against the viral infection. Eventually, through a process of recombination followed by clonal selection in multiple immune cell subtypes, the host will mount an adaptive immune response more effective against the viral infection (Figure 1). Each of the viral or host process includes many subprocesses and interactions. Dynamic host-virus interactions can be used to explain disease outcomes according to their impact on these many subprocesses. If the viruses "wins", the result will be the viral survival and replication. If the host "wins", the host will eventually control viral replication and kill the viruses. To model and analyze the complex adversarial interactions between host and pathogen, we propose to apply ontology as an integrative and systematic platform due to the solid semantic logic behind ontology and its computer-interpretable nature.

**Coronaviruses causing different host outcomes belong to different categories.**

We used ontology to represent coronavirus disease knowledge including the taxonomical structure of 9 coronaviruses in the NCBITaxon ontology, a taxonomy ontology based on the NCBI Taxonomy database[17]. As shown in Figure 2A, coronaviruses belong to subfamily Orthocoronavirinae ("true" coronaviruses) in the family Cornavirinae. SARS-Cov and SARS-Cov-2 belong to the Sarbecovirus subgenera in genus Betacoronavirus. MERS-CoV belongs to the Merbecovirus subgenera, also in genus Betacoronavirus. Three human CoV strains HKU1, OC43, and A59 belong to subgenus Embecovirus in genus Betacoronavirus. However, the two other human-infecting strains (229E and NL63) belong to genus Alphacoronvirus. Avian coronaviruses, such as IBV, are Gammacoronaviruses. As part of the NCBITaxon ontology, this Linnean hierarchy is extracted to become part of the CIDO.

**Human coronaviruses infect and transmit among mammal hosts.** We identified over 10 organisms that can serve as hosts or reservoirs for human coronaviruses. Bats were found to be the reservoir of a wide variety of coronaviruses, including SARS-CoV and MERS-CoV viruses [27]. SARS-CoV was also found in masked palm civets and a raccoon dog, and antibodies against SARS-CoV were also found in Chinese ferret badgers in a live-animal market in Shenzhen, China [28]. Our analysis found many other animals that might have served as reservoir hosts of SARS-CoV (Figure 2B). The MERS-CoV virus likely transmitted from dromedary camels [29].

These host organisms can further be classified using the NCBITaxon ontology (Figure 2B). All the host organisms belong to Boreoeutheria, a clade (magnorder) of placental mammals. Clades within Boreoeutheria include Euarchontoglires (Supraprimates, including humans and rodents) and Laurasiatheria. Laurasiatheria includes Artiodactyla (even toed ungulates, such as cattle and camels), Chiroptera (bats), and Ferae (carnivores such as canines, felines, and civets, and relatives including pangolins) (Figure 2B). The taxonomy hierarchy was also extracted from the NCBITaxon ontology and become part of the CIDO.

Our NCBITaxon-based ontology analysis of all these coronavirus hosts further shows that zoonotic human coronaviruses (including SARS-CoV, MERS-CoV, and SARS-CoV-2) are obligate mammal parasites (Figure 2B), suggesting that some molecular factor distinguishes mammals from other animals to host and transmit coronaviruses to humans. We hypothesize that zoonotic human coronaviruses (such as SARS-CoV-2) can only originate in mammals (in contrast to influenza, which can transmit to humans from avians). However, this hypothesis will need substantial additional study. A recent paper suggests snake as an intermediate host for human zoonotic coronavirus based on similarity in codon usage bias with 2019-nCoV (e.g., SARS-CoV-2) [30]. Since snakes are not mammals, this would disprove our hypothesis, but the overall pattern makes the proposed transmission via snakes an extreme claim, thus requiring extreme evidence.

**COVID-19 outcomes differ among hosts.** According to US CDC, the common symptoms of COVID-19 include fever, cough, shortness of breath or difficulty breathing, chills, repeated shaking with chills, muscle pain, headache, sore throat, and new loss of taste or smell. Figure 3A is a hierarchical display of these 9 common symptoms and their associated hierarchical phenotype representation using the Human Phenotype Ontology (HPO) [31].

Comorbidity is the presence of additional condition(s) such as a disease or phenotype that is concurrent with the coronavirus infection. Our further COVID-19 annotation and analysis show that the patients with many comorbid conditions (e.g., emphysema, chronic bronchitis, hypertension, coronary heart disease, and diabetes) are at heightened risk for severe symptoms and death. Figure 3B shows comorbidities and the COVID-19 outcome frequencies from case reports in peer-reviewed articles. To analyze the host interactions leading to these outcomes, it is very important to build data structures and data visualizations that can draw upon clinical electronic records.

**Literature mining, annotation, and analysis of host-coronavirus interactions.**

Clinical coronavirus disease outcomes result from host-coronavirus interactions. We have collected 10, 4, and 1 virulence factors from SARS-CoV, MERS-CoV, and SARS-CoV-2, respectively (Table 1, Figure 4). Table 1 shows 35 manually curated host-coronavirus protein-protein interactions (PPIs), which has been extracted from peer-reviewed articles and stored in Victors, a virulence factor database we developed[15]. These PPIs include a number of experimentally verified mechanisms. For example, all betacoronaviruses encode a spike (S) surface glycoprotein that binds to their corresponding host cell receptors and mediates viral entry to the host cells. Our annotated PPIs show that SARS-CoV uses its envelope spike (S) glycoprotein to bind to its cellular receptor, angiotensin-converting enzyme 2 (ACE2)[32]. The S1 subunit of the S protein has a receptor binding domain (RBD) that binds to ACE2. Letko et al. confirmed that human ACE2 is the receptor for the SARS-CoV-2[33]. In contrast, MERS-CoV S protein binds to dipeptidyl peptidase 4 (DPP4)[34,35]. These interactions are essential for viral entry into host cells. More specifically, the viral S protein mediates the interaction with the host-cell receptor, even when the host receptor varies between viruses. Table 1 shows many other targets or effects of potential interest.

Using Ignet, our in-house centrality- and network-based literature mining analysis program[17], we performed a preliminary gene-interaction analysis on PubMed abstracts related to coronavirus. In total 14,963 coronavirus-related PubMed abstracts were extracted. Figure 4 illustrates the resulting gene interaction network, which identifies critical genes and their interactions with host factors including IFNG, TNF, ACE2,

TMPRSS2, IL1b and IL-6. IFNG had the highest degree (number of interactions) at 27, suggesting its critical regulatory roles in the coronavirus pathogenesis. ACE2 directly interacts with viral proteins of three coronaviral strains (SARS-CoV, SARS-CoV-2, and HCoV_NL63) and interacts with host-proteins that directly interact with other two strains (HCoV_229 and MERS). ACE2 is also highly connected with other genes identified by literature mining, which all strongly suggest its critical role in the pathogenic mechanisms. It is likely that patterns of interaction with ACE2 differ among coronaviruses, leading to differences in transmission and pathogenesis.

In addition to PPIs, non-PPI mechanisms also exist. Non-PPI host factors, such as host lipids [36] and nucleotides [37], are also required for viral replication. Host lipids, such as lipid rafts and cholesterol, participate in the endocytosis process by which viruses attach to and infect cells. Many naturally derived substances, including cyclodextrin and sterols, could reduce the infectivity of many viruses, including coronaviruses, by interference with lipid-dependent attachment to human cells [38]. Nucleotide and nucleoside analogue inhibitors are chemically synthesized drugs used to treat both chronic and acute viral infections by causing premature chain termination, reduced replication fidelity, and depletion of pools of naturally occurring nucleotides[37]. For example, remdesivir is a monophosphate prodrug of an adenosine nucleoside analog that has demonstrated therapeutic efficacy in a non-human primate model of Ebola virus infection [39] and against human SARS-CoV-2 infection [40].

**Unification under an ontology-based integrative framework.**

Based on our HPI-Outcome postulates, the model given in Figure 1, and our literature mining results, we generated an integrative framework to formally specify the interactions between host and SARS-CoV-2 factors with associated disease outcomes. Figure 5 illustrates how the human host interacts with SARS-Cov2 and how such interaction could result in hypertension. As discussed above, SARS-CoV and SARS-Cov2 use their envelope spike (S) glycoprotein to bind to the same cellular receptor, ACE2, which is activated by TMPRSS2[41]. The binding of the SARS-CoV/SARS-CoV2 to human ACE2 leads to the subsequent downregulation of ACE2. Experiments showed that SARS-CoV infections and the spike protein of the SARS-CoV reduce ACE2 expression, and an injection of SARS-CoV Spike into mice worsens acute lung failure, which can be attenuated by blocking the renin-angiotensin system (RAS) pathway [32]. As we know (Figure 3), patients with hypertension are more susceptible to death and severe outcomes. This phenomenon can be explained if the down-regulation of the ACE2 due to the infection is associated with hypertension (Figure 5). Specifically, the downregulation of ACE2 would lead to excessive production of angiotensin II by ACE, which would then bind to its receptor and induce vasocontraction and hypertension[42].

Although ACE and ACE2 both act as key players in the RAS, they serve oppositional roles to angiotensin II. An angiotensin II receptor blocker drug (e.g., Azilsartan) would block the angiotensin II binding to its receptor, and thus help treat hypertension. In this way, the ontological model links host-pathogen interactions to the hypertension outcome. If a patient already has hypertension, the further increased blood pressure due to the interactions would exacerbate the underlying comorbidity, which may explain the increased death rate of patients with superimposed Covid and hypertension.

However, other pathogen factors likely play a role in the extreme pathogenicity of SARS-CoV-2. This is evidenced by the fact that despite markedly weak pathogenicity, HCoV-NL63 (a prevalent human coronavirus) also uses an S-ACE2 interaction for cellular entry[43]. However, HCoV-NL63 is considerably more benign than SARS-CoV-2. This may be a consequence of molecular differences in the S protein-ACE2 interaction between the two strains. The ontology-driven comparative analysis of CoVs provides a formal representation for both similarities and differences among the S proteins from different viruses. If the S proteins do not differ significantly, we hypothesize that some other factor (*e.g.*, a co-receptor) of SARS-CoV-2 contributes to viral binding and entry. Our recent vaccine design study[44] predicts many SARS-CoV-2 nonstructural proteins as possible adhesins, some of which may serve as a possible co-receptor, which are testable predictions.

A cytokine storm is an excessive immune response to external stimuli associated with high levels of cytokine production[45]. Cytokine storm is a major cause of acute respiratory distress syndrome (ARDS) and multiple-organ failure in severe COVID-19 patients, resulting in death within a short time[46]. IL-6 is a major cytokine involved in the cytokine storm[45]. IL-6 can be induced by the angiotensin II in human cells [47], which links the cytokine storm to the S-ACE2 binding-associated interaction network (Figure 5). Tocilizumab, a recombinant humanized anti-human IL-6 receptor monoclonal antibody, can bind to the IL-6 receptor with high affinity, and thus prevent IL-6 from binding to its native receptor and inducing downstream immune damage. It was recently reported the Tocilizumab improved the clinical outcome in severe and critical COVID-19 patients and effectively reduced mortality [48].

Figure 5 provides a logical and structured graphical summary of the knowledge introduced above. This representation covers knowledge learned from many studies. Without such an integrative representation, it is difficult to put all the pieces together. Another unique feature of the Figure 5 representation is that it is ontology-oriented, so that all the terms and relations are represented using a computer-interpretable ontological format. Figure 5 is a subset of the comprehensive host-coronavirus interaction network, which will be expanded as our understanding of host-interactions by the coronavirus grows.

To demonstrate the ontological representation, the *process 'SARS-CoV-2 S-ACE2 binding'* can be defined by two logical axioms (Figure 5):

*'has participant' some ('SARS-CoV-2 S protein' and ('has role' some 'ligand role'))*
*'has participant' some ('ACE2' and ('has role' some 'receptor role'))*

The above ontological logical axioms define the relations where a SARS-CoV-2 S-ACE2 binding process has two participants: a SARS-CoV-2 S protein that serves as a ligand role, and an ACE2 protein that serves as a receptor role.

In addition, we can define a logical axiom between 'SARS-CoV-2 S-ACE2 binding' and TMPRSS2 that defines the role of TMPRSS2 to activate the SARS-CoV-2 S-ACE2 binding process:

*TMPRSS2: activates some 'SARS-CoV-2 S-ACE2 binding'*

A major difference between ontology and taxonomy is that a taxonomy is built only based on the is_a relation. Beside the representation of such is_a relation, ontology can also provide additional relations between entities under different hierarchies, such as the above 'activates' relation that links the TMPRSS2 protein in the protein hierarchy and the SARS-CoV-2 S-ACE2 binding under the interaction process hierarchy.

Our ontology-based integrative framework will also cover other parts of host-coronavirus interactions and their relations to the disease outcomes. Zoonotic human coronaviruses appear to delay active innate responses to viral infection. The interferon (IFN) family of cytokines, including IFN-α, IFN-β and IFN-γ, tigger the early innate response against viral pathogens. However, SARS-CoV and MERS-CoV are able to delay IFN induction and dysregulate IFN-stimulated gene (ISG) effector functions in primary human airway epithelial cells or in cultured cells (altered histone modification is a proposed molecular mechanism)[49,50]. Inteferons, in turn, are able to initiate the expression of hundreds of ISGs that have antiviral and immune regulatory functions [49,51]. The excessive production of pro-inflammatory cytokines and chemokines, esp. IL-1b, IL-8, IL-6, CXCL10, and CCL2, were detected in SARS patients [52-54], indicating the initiation of a cytokine storm, which is eventually associated with the progression to ARDS and death. It is obvious that the timing here is critical. The early IFN response would be beneficial. However, coronaviruses, especially SARS-CoV-2, manipulate the immune system to delay such a response allowing widespread infection prior to an onset of symptoms. The understanding of such dynamics helps our rational design of proper treatments and the timing of the treatments.

It appears that similar to SARS-CoV and MERS-CoV, SARS-CoV-2 also affects the IFN response [51,55]. SARS-CoV-2 even displayed a higher sensitivity to IFN-I in vitro compared with SARS-CoV-1 in infected Vero cells [56]. The difference in terms of IFN-I activation between SARS-CoV and SAR-CoV-2 is likely due to the difference in the sequences for two proteins (Orf 3b and Orf 6) of these two human coronaviruses [56]. Considering that different human coronaviruses induce varied phenotype outcomes, it is important to include more coronaviruses in the systematic comparison and identify their differential action on IFN production and on ISGs, which likely plays a critical role in the delayed host recognition and responses to the infection, supporting the survival and replication of the viruses inside the host. Our proposed HPI postulates combined with our structured ontology support an systems-level investigation into these mechanisms and into potential avenues of prophylaxis or treatment.

The second key component of the host response to coronavirus interaction is the activation of an adaptive host response, which is also key to long term immunity. The severe SARS disease is also related to the lack of or delayed activation of adaptive immunity [42]. The SARS patients with severe outcomes maintained high levels of many innate immune system-linked cytokines (e.g., CXCL10, CCL2) and ISG-encoded proteins with low levels of spike-specific antibodies [57]. Conversely, surviving patients had higher levels of spike-specific antibodies. These findings suggest that the prompt formation of adaptive immunity is important to reduce disease severity and improve outcomes. The adaptive immune response can be induced by effective vaccines against COVID-19. Both SARS-CoV and MERS-CoV S and S1 have been used as vaccine antigens/immunogens due to their ability to induce neutralizing antibodies that prevent host cell attachment and infection [44,58]. Numerous SARS-CoV-2 vaccine trials are underway [44], with the hope that a safe and effective vaccine(s) will be ready in the near future to control the COVID-19 pandemic.

**Computational prediction of new host-coronavirus PPIs and tissue responses**

Our previous Victors PPI predictions are based on the analyses of protein domains and phylogeny[15]. By nature, each PPI is a domain-domain interaction. Therefore, we can identify the domains of virus proteins, and based on known interactions with domains of host proteins, predict how the virus proteins interact with host proteins. Using the known PPIs as baits, we developed a domain-inferred PPI prediction strategy to predict more PPIs, and how these PPIs could possibly affect organs and tissues. Our prediction found that SARS-CoV-2 could induce 3,111 human-virus interactions with 133 human proteins. Many PPIs are tissue-specific, and the tissue-specific information can be identified from the Human Protein Atlas [59]. Using our predicted PPIs, SARS-CoV-2 was predicted to affect many tissues such as skin, liver, intestine, blood, glands and reproductive tissues (Figure 6).

In addition to computational methods, the prediction described above also depends on prior knowledge of the pre-identified PPIs and tissue-specific protein expression profiles. Currently, these pre-identified PPIs and tissue-preferred gene profiles are stored in tables or databases that do not have a scientifically well-motivated associated ontology. Therefore, the use of this prior knowledge is labor intensive and difficult to automate. We can ontologically represent the domains from viral and host proteins, individual domain-domain interactions, and the tissue preference of each protein and domain. We expect that such a systematic ontology-based representation of this prior knowledge would facilitate the usage of both domain-domain interactions and tissue specificity studies, supporting additional interaction predictions and analysis.

**Discussion**

The contribution of this article is manifold. First, a set of host-pathogen interaction (HPI)-outcome postulates are proposed to provide a framework for deeper understanding of host-pathogen interactions and their causal role in disease outcomes. The HPI-outcome postulates can be applied to fundamentally understand interactions between hosts (including humans) and coronaviruses (including SARS-CoV-2). Second, to use the HPI-outcome postulates, we proposed an integrative ontology-based framework to systematically model, represent, and analyze human-coronavirus interaction (HCI)-outcome relations, including coronavirus and host phylogenies, and literature mined and manually annotated host-coronavirus interactions, drugs, vaccines, and their intertwined relations. Third, examples such as a potential mechanistic link between the S-ACE2 interaction and blood pressure comorbidities illustrate applications of our integrative ontology-based framework. We demonstrate a structured, hierarchical representation of coronaviruses, hosts, phenotypes, and HCI interactions in terms of viral pathogenesis and host innate and adaptive responses. We showed that based on prior knowledge, we could develop computational predictions to predict new host-coronavirus interactions (HCIs) and extrapolate to hypothetical affects on different tissues and organs.

We envision that our proposed integrative ontology-based framework could serve as a foundation for ongoing and deeper studies of the COVID-19 disease. We have initiated the development of the community-driven Coronavirus Infectious Disease Ontology (CIDO), which can serve as a logic framework for the systematic representation of the HCIs, disease outcomes, and the relations between the HCIs and disease outcomes[14]. CIDO is an interoperable with other ontologies, such as the community-based Ontology of Host-Pathogen Interactions (OHPI)[15]. The community-based CIDO can serve as an ontology platform for describing host-coronavirus interactions (HCIs) built from these HPI-postulates.

The intergrative ontological modeling can also support different applications such as rational drug repurposing design and vaccine development. As shown in Figure 5, our ontological representation provides several paths for rational drug repurposing design for COVID-19 treatment. Camostat mesylate (a serine protease inhibitor) is capable of inhibiting the function of TMPRSS2[41], which is crucial for the effective S-ACE2 binding TMPRSS2[41]. Angiotensin II receptor blockers can be used to block the activation of the angiotensin II [60], therefore blocking the vacoconstriction and following hypertension. Tocilizumab is able to inhibit IL-6 activity[48] so block the formation of cytokine storm. SARS-CoV-2 appears to use many pathways to cause different phenotypes. It is reasonable to hypothesize that the drugs blocking early stages of the disease pathways would be more effective than the later starges of the pathways. It is also important to consider drugs that block the paths leading to severe disease outcomes such as hypertension and cytokine storm. In addition, we also need to consider the side effects of the drugs being used for the treatment. The integrative ontology framework provides a foundation for these rational design. Ontology-based software tools and algorithms can also be developed to support intelligence queries[61,62] and enhance computational predictions[63-65], especially when the computer-intepretable ontology knowledge base system becomes big and complex.

The CIDO ontology-based evidence-driven framework can provide a fundamental structure analgous to Mendeleev's periodic table of chemical elements. Based on the 64 chemical elements known, Mendeleev generated a periodic table to position these elements and predicted the deeper, underlying structure giving rise to their chemical properties. Similarly, we can accumulate all the existing evidence of HCIs and put these interactions into a HCI framework to identify periodic patterns, and use these patterns to predict strategies for drugs and vaccines. These patterns are represented through an ontology, which provides a periodic table-like framework. CIDO aims to become such an ontology. Our future plan includes more logical and thorough representation and analysis of the host-coronavirus molecular and cellular interactions, and development of new ontology-based algorithms and tools to support translational applications.

**Acknowledgements**

This project is supported by NIH grants U24CA210967 and P30ES017885 (to GSO); the non-profit Central Research Institute Fund of Chinese Academy of Medical Sciences 2019PT320003 (to HY); and University of Michigan Medical School Global Reach award (to YH).

**FIGURE LEGENDS**

**Figure 1. Model of host-coronavirus interactions and their associated disease outcomes.** The viruses enter into, survive in, and replicate in host cells. After initial naïve acceptance of viral entry without triggering an immune reaction (a naïve response), the host initiates active innate and adaptive responses.

**Figure 2**. **Ontological representation of coronaviruses, hosts, and the coronavirus-host associations.** (A) The taxonomical hierarchy of human coronaviruses and a few other coronaviruses as outgroups. (B) The taxonomical hierarchy of reported hosts of coronaviruses. The links represent associations between the coronaviruses and established hosts.

**Figure 3**. **Human phenotypes associated with COVID-19.** (A) Human phenotypes commonly seen in COVID-19 patients. (B) Hierarchical representative of comorbidity phenotypes and associated phenotype frequency in mild and severe COVID-19 patients. For example, (0.14, 0.30) in (B) indicates that superimposed hypertension is found in 14% and 30% of mild symptom and severe symptom patients, respectively. The results were summarized from reported literature [66,67].

**Figure 4. Gene-gene interaction network by Ignet on the PubMed abstracts related to coronavirus.** The figure was constructed based on expert curated (left non-gray colored) and literature mined gene interaction by Ignet from 14,963 abstracts, retrieved by a PubMed search of "coronavirus". Node (gene) and font size correspond to the number of connections to other nodes, a.k.a. degree. Edge (interaction) thickness represents the relative importance of each connection in the network, measured by edge betweenness centrality. This network includes 131 nodes with 163 edges.

**Figure 5. Model linking host-coronavirus interactions to outcomes.** This model includes different components that integrate into the systematic framework, which illustrates a potential link between SARS-CoV-2 infection and hypertension severity (and cytokine storm).

**Figure 6. Predicted protein-protein interactions (PPIs) and their affected tissues.** PPIs are predicted based on domain-domain interactions and sequence similarity. We predicted 3,111 human-virus interactions involving 133 human proteins. The nodes in the middle are virus proteins and nodes in the outer circle are human proteins. The interacting human proteins are annotated with tissue-specific expression profiles from the Human Protein Atlas. Proteins enriched in different tissues are marked with different colors.

**Table 1. Known human-coronavirus protein-protein interactions (PPIs)**

| Viral protein (gene) | Human protein (gene) | Description | References |
|---|---|---|---|
| **Human coronavirus 229E** | | | |
| Spike glycoprotein (S) | Aminopeptidase N (ANPEP) | viral attachment and entry | 1350662 |
| | Tubulin beta-2A chain (TUBB2A) | viral transportation, localization, and assembly | 27479465 |
| | Tubulin alpha-4A chain (TUBA4A) | | |
| | Tubulin beta-4A chain (TUBB4A) | | |
| | Tubulin beta-6 chain (TUBB6) | | |
| **Severe acute respiratory syndrome coronavirus** | | | |
| Spike glycoprotein (S) | Angiotensin-converting enzyme 2 (ACE2) | viral attachment and entry | 15897467 |
| | C-type lectin domain family 4 member M (CLEC4M) | viral attachment and entry | 15496474 |
| | Vimentin (VIM) | viral attachment and entry | 26801988 |
| Envelope small membrane protein (E) | Bcl-2-like protein 1 (BCL2L1) | induce T-cell apoptosis | 16048439 |
| Nucleoprotein (N) | Heterogeneous nuclear ribonucleoprotein A1 (HNRNPA1) | viral RNA synthesis | 15862300 |
| | Peptidyl-prolyl cis-trans isomerase A (PPIA) | viral invasion or replication | 15688292 |
| Protein X1 (sars3a) | Caveolin-1 (CAV1) | viral assembly, uptake, and trafficking | 17947532 |
| Protein X2 (orf3b) | Runt-related transcription factor 1, isoform b (RUNX1b) | viral pathogenesis | 22253733 |
| Protein X3 (orf6) | Importin subunit alpha-1 (KPNA2) | evade host immune response | 17596301 |
| Protein 7a (7a) | Small glutamine-rich tetratricopeptide repeat-containing protein alpha (SGTA) | viral assembly and release | 16580632 |
| | Integrin alpha-L (ITGAL) | viral attachment and entry | 18020948 |
| | Bcl-2-like protein 1 (BCL2L1) | induce apoptosis | 17428862 |
| | Bcl-2-like protein 2 (BCL2L2) | induce apoptosis | 17428862 |
| | Induced myeloid leukemia cell differentiation protein Mcl-1 (MCL1) | induce apoptosis | 17428862 |
| Protein 9b (9b) | Exportin-1 (XPO1) | induce apoptosis | 21637748 |
| Nsp3-pp1a/pp1ab (nsp3) | Interferon regulatory factor 3 (IRF3) | evade host immune response | 17761676 |
| Nsp5-pp1a/pp1ab (nsp5) | V-type proton ATPase subunit G 1 (ATP6V1G1) | viral pathogenesis | 16226257 |
| Nsp10-pp1a/pp1ab (nsp10) | Transcription factor BTF3 (BTF3) | viral replication | 16157265 |
| | Cyclic AMP-dependent transcription factor ATF-5 (ATF-5) | viral replication | 16157265 |

|  | NADH-ubiquinone oxidoreductase chain 4L (MT-ND4L) | viral replication | 16157265 |
|---|---|---|---|
|  | Cytochrome c oxidase subunit 2 (MT-CO2) | viral replication | 16157265 |
| **Human coronavirus NL63** | | | |
| Spike glycoprotein (S) | Angiotensin-converting enzyme 2 (ACE2) | viral attachment and entry | 15897467 |
|  | Tubulin beta-2A chain (TUBB2A) | viral transportation, localization, and assembly | 27479465 |
| **Middle East respiratory syndrome-related coronavirus** | | | |
| Spike glycoprotein (S) | Dipeptidyl peptidase 4 (DPP4) | viral attachment and entry | 23486063 |
| Envelope small membrane protein (E) | TNF receptor-associated factor 3 (TRAF3) | evade host immune response | 27094905 |
| Non-structural protein ORF4a (orf4a) | Interferon-inducible double-stranded RNA-dependent protein kinase activator A (PRKRA) | evade host immune response | 24522921 |
| Non-structural protein ORF4b (orf4b) | Inhibitor of nuclear factor kappa-B kinase subunit epsilon (IKBKE) | evade host immune response | 26631542 |
|  | Interferon-induced helicase C domain-containing protein 1 (IFIH1) | evade host immune response | 26631542 |
|  | Serine/threonine-protein kinase TBK1 (TBK1) | evade host immune response | 26631542 |
| **Novel coronavirus** | | | |
| Spike glycoprotein (S) | Angiotensin-converting enzyme 2 (ACE2) | viral attachment and entry | 31996437 |

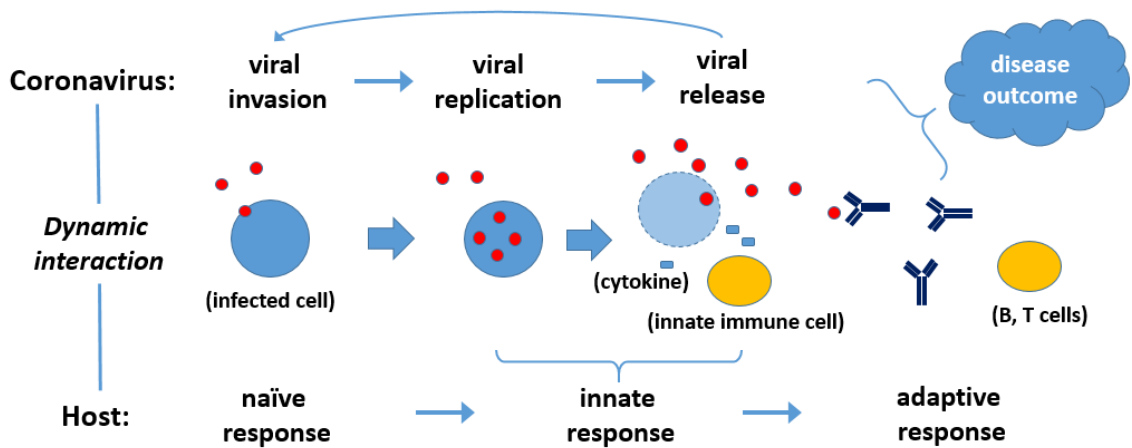

**Figure 1. Model of host-coronavirus interactions and their associated disease outcomes.** The viruses enter into, survive in, and replicate in host cells. After initial naïve acceptance of viral entry without triggering an immune reaction (a naïve response), the host initiates active innate and adaptive responses.

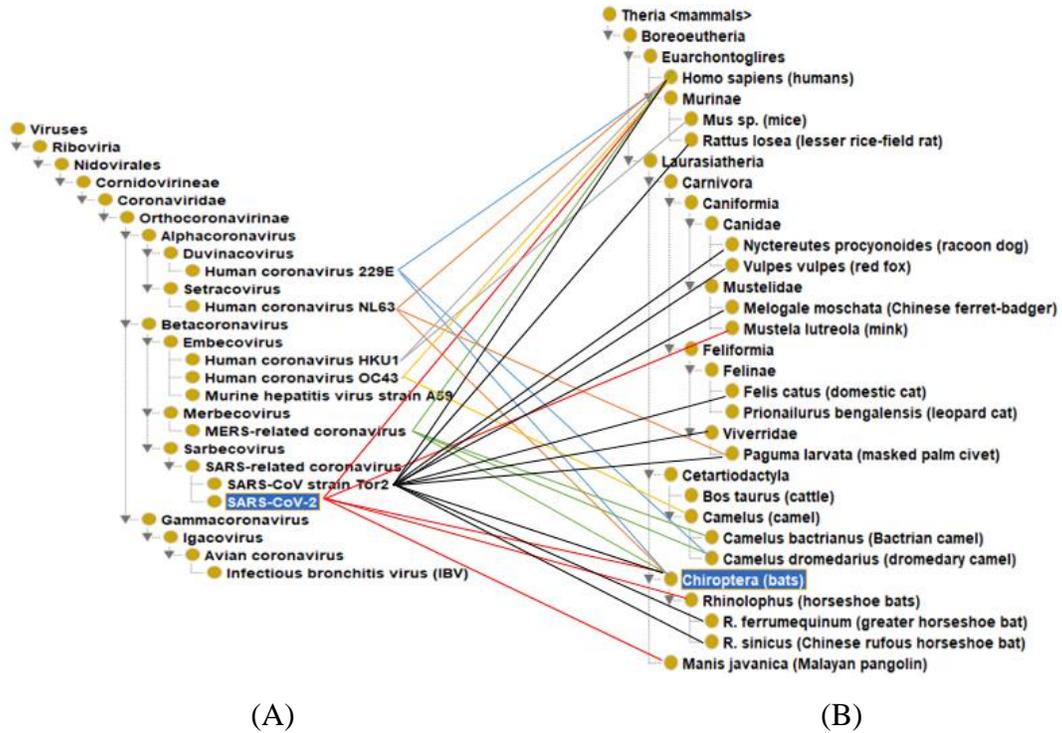

(A)                                   (B)

**Figure 2**. **Ontological representation of coronaviruses, hosts, and the coronavirus-host associations.** (A) The taxonomical hierarchy of human coronaviruses and a few other coronaviruses as outgroups. (B) The taxonomical hierarchy of reported hosts of coronaviruses. The links represent associations between the coronaviruses and established hosts.

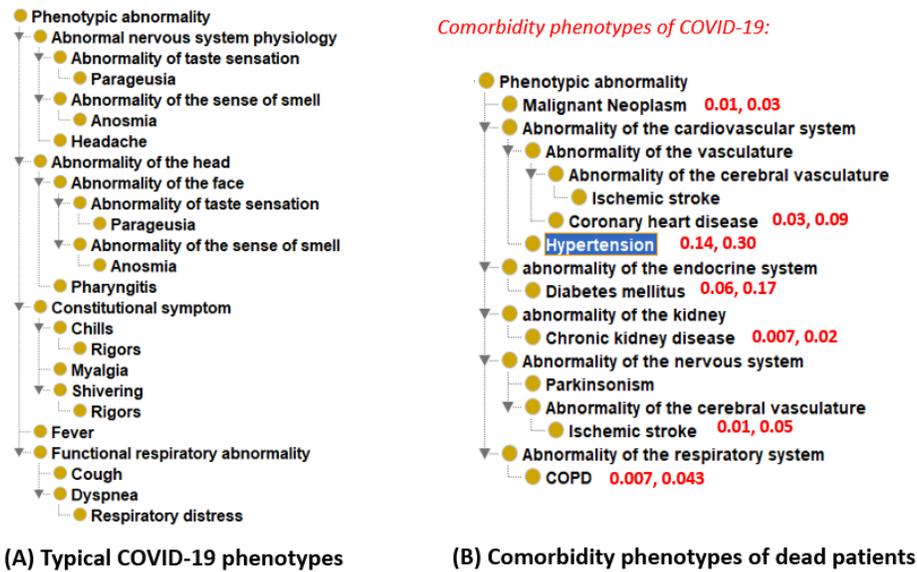

**Figure 3**. **Human phenotypes associated with COVID-19.** (A) Human phenotypes commonly seen in COVID-19 patients. (B) Hierarchical representative of comorbidity phenotypes and associated phenotype frequency in mild and severe COVID-19 patients. For example, (0.14, 0.30) in (B) indicates that superimposed hypertension is found in 14% and 30% of mild symptom and severe symptom patients, respectively. The results were summarized from reported literature [66,67].

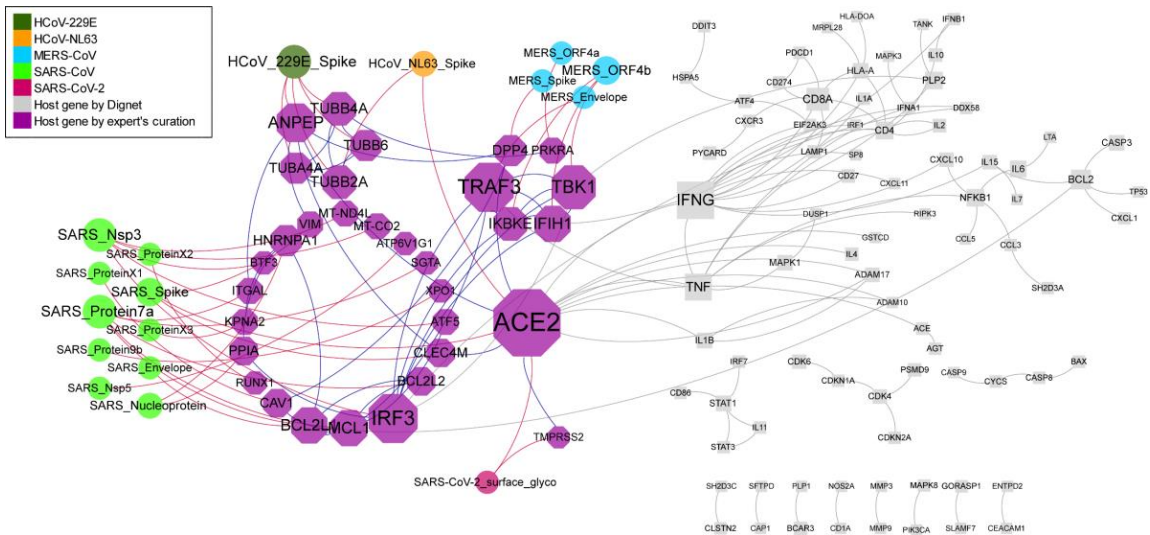

**Figure 4. Gene-gene interaction network by Ignet on the PubMed abstracts related to coronavirus.** The figure was constructed based on expert curated (left non-gray colored) and literature mined gene interaction by Ignet from 14,963 abstracts, retrieved by a PubMed search of "coronavirus". Node (gene) and font size correspond to the number of connections to other nodes, a.k.a. degree. Edge (interaction) thickness represents the relative importance of each connection in the network, measured by edge betweenness centrality. This network includes 131 nodes with 163 edges.

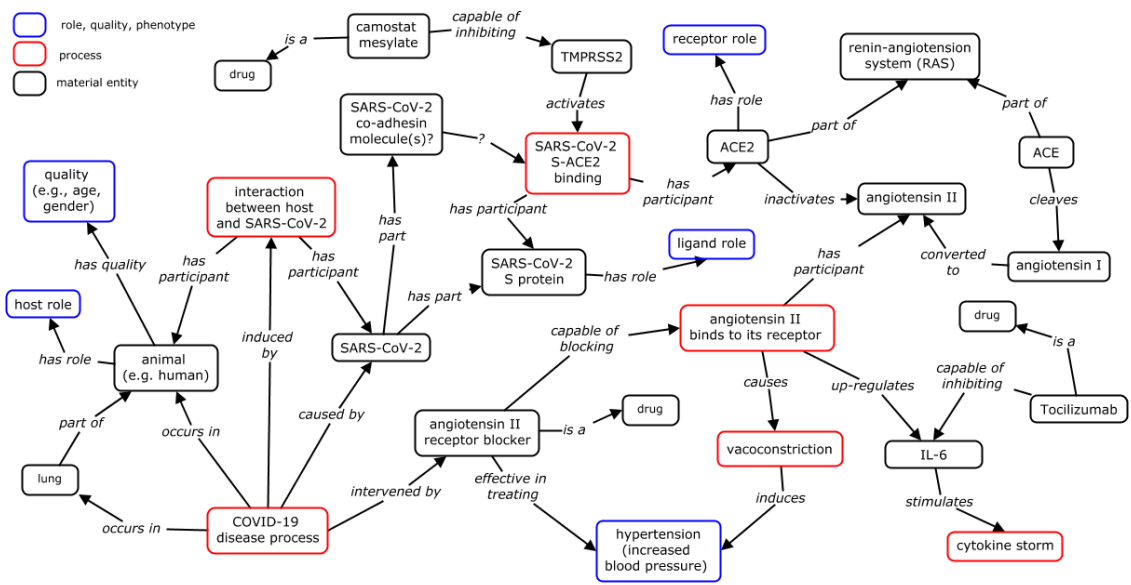

**Figure 5. Model linking host-coronavirus interactions to outcomes.** This model includes different components that integrate into the systematic framework, which illustrates a potential link between SARS-CoV-2 infection and hypertension severity (and cytokine storm).

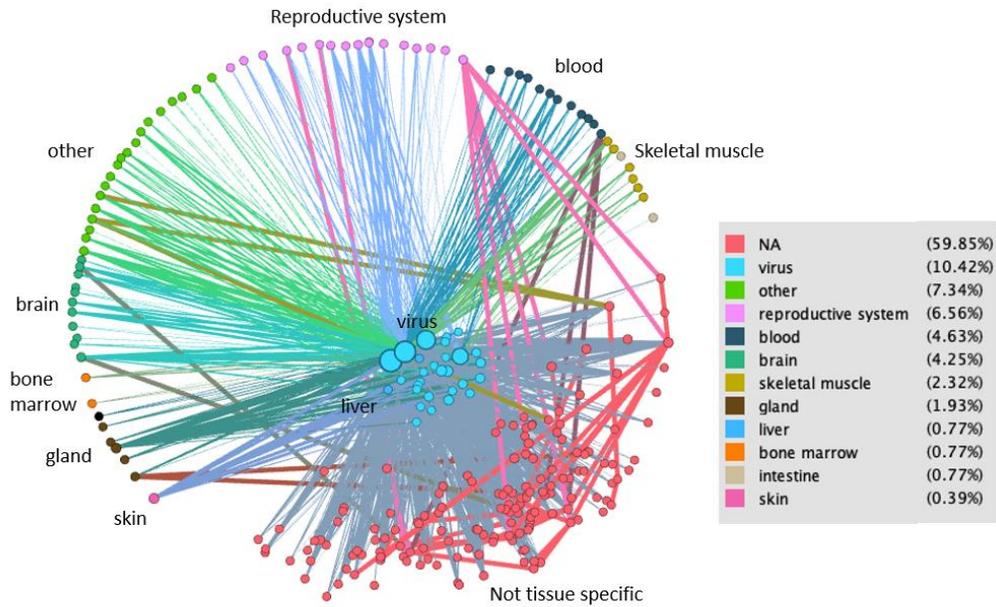

**Figure 6. Predicted protein-protein interactions (PPIs) and their affected tissues.**
PPIs are predicted based on domain-domain interactions and sequence similarity. We predicted 3,111 human-virus interactions involving 133 human proteins. The nodes in the middle are virus proteins and nodes in the outer circle are human proteins. The interacting human proteins are annotated with tissue-specific expression profiles from the Human Protein Atlas. Proteins enriched in different tissues are marked with different colors.